\documentstyle[emulapj]{article}

\font\tenbg=cmmib10 at 10pt

\def \rvecmu{{\hbox{\tenbg\char'026}}}

\begin{document}

\title{Three Dimensional MHD Simulations of Accretion to
an Inclined Rotator:  The ``Cubed Sphere"
Method}

\author{A.V.~Koldoba}
\affil{Institute of Mathematical
Modelling,
   Russian Academy of Sciences, Moscow,
Russia; koldoba@spp.keldysh.ru}

\author{M.M.~Romanova}
\affil{Department of Astronomy,
Cornell University, Ithaca, NY
14853-6801;
romanova@astro.cornell.edu}

\author{G.V.~Ustyugova}
\affil{Keldysh Institute of Applied
Mathematics,
   Russian Academy of Sciences, Moscow,
Russia; ustyugg@spp.Keldysh.ru}

\author{R.V.E.~Lovelace}
\affil{Department of Astronomy,
Cornell University, Ithaca, NY
14853-6801; RVL1@cornell.edu }

\medskip

\begin{abstract}

    We describe a three-dimensional, Godunov-type
numerical magnetohydrodynamics (MHD)  method  designed for
studying disk accretion to a rotating magnetized
star in the general case where the star's rotation
axis, its magnetic moment, and the normal
to the disk all have different directions.
     The equations of ideal  MHD are
written in a reference frame
rotating with the star with the $z-$axis
aligned with the star's rotation axis.
    The numerical method uses
a ``cubed sphere" coordinate
system which has advantages of Cartesian and
spherical coordinate systems but does
not have  the singular axis of
the spherical system.
     The grid is formed by
a sequence of concentric
  spheres of radii $R_j \propto q^j$
with $j=1..N_R$ and $q={\rm const}>1$.
     The grid on the surface of the
sphere consists of six sectors with the grid on
each sector  topologically equivalent to the
equidistant grid on the face of a
cube.
     The magnetic field
is written as a dipole component plus
deviations, and only  the deviations
are calculated.
     Simulation results are discussed for
the funnel flows (FF) to a star with
dipole moment $\rvecmu$  at an
angle $\Theta=30^\circ$ to the star's rotation
axis ${\bf \Omega}$ which is aligned with the normal to the
disk.
    Results are given for different grids
($N_R \times N^2 $), from
$26\times 15^2$ (coarsest)
to $50\times 29^2$ (finest).
   We observe that the
qualitative features of the accretion flows
are very similar for the different grids, but
the coarser grid is affected by numerical
viscosity.
    We compare our  $3D$ results for $\Theta=0$ with
axisymmetric ($2D$), spherical coordinate
system simulations of funnel flows
  (Romanova {\it et al.} 2002).
   Two important new 3D features are found in these
simulations: 
(1)  The funnel
flow to the stellar surface  is mainly in  two streams
which approach the star from opposite directions.
  (2) In the $x-z$ cross section of the flow containing
$\rvecmu$ and ${\bf \Omega}$, the funnel flow often takes
the longer of the two possible paths along magnetic
field lines to the surface of the star.  
     A subsequent paper will give a detailed 
description of the method and
results on $3D$ funnel flows at different 
inclination angles $\Theta$.

\end{abstract}


\section{Introduction}

   There are many X-ray binary systems
consisting of a rotating neutron star
with dipole magnetic field which
accretes matter from the other star.
    Commonly, the accreting matter forms
a disk as it approaches the neutron star.
    The neutron star's magnetic dipole
axis $\rvecmu$ is not aligned with its rotation
axis ${\bf \Omega}$ in the observed pulsating sources.
    Furthermore, in general $\rvecmu$ and ${\bf \Omega}$
are not aligned with the normal to the disk.
That is, the dipole moment is inclined
relative to the disk. In some systems,
such as Her X-1  this inclination is
evident from the observations (e.g.,
Tr$\ddot{\rm u}$mper {\it et al.} 1986).
     Models of magnetohydrodynamic (MHD) disk
accretion to a rotating star with
an aligned dipole magnetic field
have been discussed by a number of authors
(e.g., Ghosh \& Lamb 1979;
Camenzind 1990; K\"onigl 1991;
Lovelace, Romanova, \&
Bisnovatyi-Kogan 1995, 1999;
Ostriker \& Shu 1995;
Li \& Wilson 1999;  Koldoba {\it et al.} 2002).
         Analytical investigations of
three dimensional (3D) accretion to an
inclined magnetic rotator were done by
Arons and Lea (1976a,b), Lipunov
(1978a,b), Sharlemann (1978), Aly
(1980), Lai (1999), and by
Terquem and Papaloizou (2000).
  However, the essential aspects of this problem are
three-dimensional, and  these are not
amenable to an analytic approach.

    In this paper we describe a method
  developed specifically for studying
magnetohydrodynamic accretion to non-aligned
magnetic rotators.
   The method is based on
the ``cubed sphere"  coordinate system
proposed by  Sadourny (1972)
and   discussed
by Ronchi, Iacono, \& Paolucci (1996)
(hereafter RIP96).
     We solve the equations of ideal
magnetohydrodynamics (MHD) in
a reference frame rotating with
the star using ``cubed
sphere'' coordinates and using a Godunov-type
method.
    In contrast with RIP96,  our simulations are
in $3D$ rather than on
the surface of the sphere, and furthermore
we solve the equations of MHD.

     An important question is what
grid resolutions are necessary for
obtaining valid results for the problem
of   MHD disk accretion to a rotating star
with magnetic moment mis-aligned with
the star's rotation axis?
     Because  $3D$ MHD simulations
are very  time-consuming,
it is important to find an optimal grid resolution.
   Here, we discuss results from
simulations for the
case where the star's magnetic
moment is inclined at an
angle $\Theta=30^\circ$ to the star's rotation
axis which is aligned with the normal to the
disk.
     In \S 2 we describe our simulation methods.
In \S 3 we summarize results of our $3D$ simulations
of funnel flows to a rotating mis-aligned dipole,
and we compare $3D$ simulation results for an aligned
dipole with our earlier axisymmetric simulation
results (Romanova {\it et al.} 2002).
   Conclusions of this work are given in \S  4.

\begin{figure*}[t]
\centering
 \figcaption{
Illustration of the ``cubed sphere''
grid which consists of six sectors.
  The $+x$ sector is
omitted in order to
show the inner structure of the grid.
   The grid is inhomogeneous in the $R-$direction
in such a way that  the cells are
roughly cubical independent of $R$.
} 
\end{figure*}

\section{Simulation Method}

      We consider a rotating magnetized star
surrounded  by an accretion disk and its corona.
     This problem is difficult to treat
numerically because the  magnetic field varies
strongly with distance from the star
($\sim 1/R^3$), and it is rapidly varying
in the laboratory inertial reference frame.
To minimize errors in calculating
the magnetic  force, the magnetic
field $\bf B$ is decomposed into 
the ``main'' dipole component of the
 star,  ${\bf B}_0$, and the
component, ${\bf B}_1$, induced by 
currents in the disk and in the corona. 
  Because
${\bf \nabla}\times{\bf B}_0=0$, the magnetic force
is $({\bf J}\times{\bf B})/c =
 (\nabla\times {\bf B_1})\times
({\bf B}_0 + {\bf B}_1)/4\pi$, which does not involve
terms  $ {\cal O}({\bf B}_0^2)$ (Tanaka 1994; Powell et al.
 1999).

Another difficulty in this problem is that the dipole
moment changes with time. It rotates with angular
velocity ${\bf \Omega}$ so that the ``main" field ${\bf B}_0$
also changes with time. 
   Consequently in
the induction equation there is a
large  term involving ${\bf B}_0$.
To overcome this difficulty we use a coordinate system
 rotating with angular
velocity ${\bf \Omega}$, in which the
 magnetic moment of the star ${\rvecmu}$ and the ``main"
field ${\bf B}_0$ do not depend on time.

\subsection{MHD Equations in the
Rotating Frame}

Let $\bf \Omega$ be angular velocity
of rotation of the star.
In the laboratory (inertial)
reference frame the MHD
equations are
$
{{ D} \rho}/{{ D} t} +
\rho {\bf
\nabla} \cdot {\bf u} = 0$,
 $ \rho {{ D} {\bf
u}}/{{ D} t} =- {\bf
\nabla} p +
 ({\bf \nabla \times
B})\times {\bf  B}/4\pi +
\rho {\bf g},
$
$
{{ D} S}/{{ D} t}=0~,
$
$
{{ D} {\bf B}}/{{ D} t} =
   ({\bf B}\cdot
{\bf\nabla}){\bf u} -
{\bf B} ({\bf\nabla} \cdot {\bf u})~.
$
   Here, $\bf u$ is bulk velocity of the plasma,
$S$  the specific entropy,
$\bf B$  the magnetic field, and ${
D}/{D t} = {\partial}/{\partial
t} + {\bf u}\cdot{\bf \nabla} $
the convective derivative in
the inertial frame.

     The relations between the variables
in the inertial  frame
and the reference frame rotating
at the rate ${\bf \Omega}$  for a given
fluid particle are
$ {\bf u} = {\bf v} + {\bf
\Omega}\times{\bf R}$,
${\bf B} = {\bf B}$,
where ${\bf v}$ is velocity of plasma
in the rotating frame.
     The convective
derivatives of scalar variables $f$
do not change, ${{ D} f}/{{ D}
t} = { df}/{d t}$, where $d/{d t}
= {\partial}/{\partial t} + ({\bf
v}\cdot{\bf\nabla})$ is the convective
derivative in the rotating frame.
     However, convective
derivatives of vector variables
transform as following:
   ${{D} {\bf F}}/{{ D} t} = {
d {\bf F}}/{d t} + {\bf
\Omega}\times{\bf F}$.

    Because ${\bf \nabla}\cdot ({\bf
\Omega}\times {\bf R}) = 0$, the
  continuity equation is
$
{{ d} \rho}/{{ d} t} +
\rho {\bf
\nabla} \cdot {\bf v} = 0 ~.
$
In the Euler equation  in
the rotating frame, there
are two inertial terms  added to the right hand
side, the Coriolis force
and the centrifugal force,
$
\rho {d{\bf v}}/{dt} = -{\bf
\nabla} p + ({\bf \nabla \times
B})\times{\bf B}/4\pi + \rho{\bf g}
   + 2\rho~{\bf
v}\times{\bf\Omega}
-\rho~{\bf\Omega}\times({\bf\Omega}\times{\bf
R}).
$
In the induction equation
$
   ({\bf B}\cdot{\bf\nabla}){\bf u} =
({\bf B}\cdot{\bf \nabla}){\bf v}+
   ({\bf B}\cdot
{\nabla})({\bf\Omega}\times{\bf R})
=({\bf B}\cdot{\bf\nabla}){\bf v} +
{\bf\Omega}\times{\bf B}~.
$
Therefore in the rotating reference frame,
the induction equation has the
same form,
$
{d{\bf B}}/{dt}=
({\bf B}\cdot{\bf\nabla}){\bf v}
   -{\bf B} ({\bf \nabla}\cdot {\bf v}).
$

     Putting the equations in Eulerian form
in the rotating frame, one finds
that the new terms are the Coriolis
and centrifugal forces in the Euler
equation.
    The full set of equations is
$
{{\partial \rho}/{\partial t}}
+ {\bf {\nabla}}\cdot (\rho{\bf v}) = 0,
$
$
{\partial (\rho{\bf  v})}/{\partial
t} + {\bf {\nabla}}\cdot {T} =
\rho {\bf g} + 2\rho ~{\bf
v}\times{\bf \Omega} - \rho~ {\bf
\Omega}\times ({\bf\Omega}\times{\bf
R}),
$
$
{\partial (\rho S)}/{\partial t} +
{\bf {\nabla}}\cdot (\rho S {\bf v}) = 0~,
$
$
{\partial {\bf B}}/{\partial t} =
{\bf \nabla \times} ({\bf v}\times{\bf B}),
$
where $T$ is the stress tensor with components
$T_{ik} \equiv  p\delta_{ik} +\rho v_i v_k+
({B^2}\delta_{ik}/2-B_iB_k)/4\pi$. 
   We do not include shocks in the present
work as implied by the entropy 
conservation equation.

\subsection{The Grid}

    The three dimensional grid consists of a
set of concentric spheres of radii $R_j$ with
$j=1..N_R$.
       The distribution of $R_j$
is chosen to be inhomogeneous
with $R_j= R_* q^{j-1}$
where $q=$const. and $R_*$ is the radius of the
numerical star.
    This choice implies
$\Delta R_j/R_j =q-1$.
    The grid on the surface of the
sphere consists of six sectors with the grid on
each sector  topologically equivalent to the
equidistant grid on the face of a
cube.
    In each sector the grid of $N\times N$ cells
is formed by the arcs of great
circles separated by equal angles.
    For example, the $+x$ sector is formed
by the arcs of great circles going through
the $y$ and $z-$axes.

\begin{figure*}[t]
\centering
\figcaption{
Results of 3D MHD simulations of disk
accretion to a rotating star with
magnetic moment $\rvecmu$ tilted 
$\Theta=30^\circ$ to the rotation
axis ${\bf \Omega}$.
   Two grids are shown,
$N_R\times N^2 =50\times
29^2$ (the left-hand panels) and
$26\times15^2$ (the right-hand panels)
after  $t= 1.5 P_0$, where
$P_0$ is the period of rotation of the disk at $R_0$.
The shading or color represents  the density
while the lines are magnetic field lines.
  The $x-z$ cross section  is the plane
containing ${\bf \Omega}$ and $\rvecmu$.
The $x-y$ plane is the midplane of the disk at large
distances from the star.
   Note that $N_R$ is the number of grid cells
in the radial direction while
  $N\times N$ is the number of cells on the
surfaces of each of the six sectors of  
the ``cubed sphere.''
} 
\end{figure*}

\begin{figure*}[t]
\centering
\figcaption{
Three dimensional view of the funnel flow
to a rotating star with $\bf{\Omega}$
parallel to the $z-$axis and with the dipole moment $\rvecmu$
tilted $\Theta=30^\circ$ away from the $z-$axis in
the $x-z$ plane.  The grid was 
$N_R \times N^2=50\times 29^2$ and the flow is
shown  at $t=2.5P_0$, where $P_0$ is the period
of rotation of the disk at $R_0$.
    The thick arrow represents the star's magnetic
moment.  The red lines represent magnetic field lines.
The yellow spiral line is a streamline.
  The nested blue and yellow lines are isodensity lines
in the $(x,z)$ plane.
    The central  sphere represents
the  star which has a radius $R_*=0.35R_0$.
   The funnel flow close to the star is seen
to be in two streams which approach the star
from opposite directions.
    The density of the green surface shown 
is $ 0.35 \rho_0$, where $\rho_0$
is defined in \S 2.2.
} 
\end{figure*}

       This three-dimensional grid gives
  high spatial
resolution close to the star
which is important to our study of
accretion to a rotating star with
dipole magnetic field.
     Because $\Delta R_j/R_j=$const.,
the shape of the volume
elements is approximately cubical independent
of $R$.

      The present study uses the scalings introduced
by Romanova {\it et al.} (2002) where the inner
radius of the disk and the beginning of the funnel
flow (in the aligned, axisymmetric
case) is at a radial distance $R_0$.
    Distances are measured in units of $R_0$,
velocities in units of $v_0 \equiv\sqrt{GM/R_0 }$, and
time in units of $P_0 \equiv 2\pi R_0/v_0$.
    With $B_0$ the magnetic field at the
surface of the star, we can define a
reference density as $\rho_0 \equiv B_0^2/v_0^2$
and a reference value for the magnetic moment,
$\mu_0 \equiv B_0r_0^3$.
     In this study the dimensionless radius
of the numerical star is $R_{\rm min}/R_0 =0.35$, and
the outer radius of the simulation volume is
$R_{\rm max}/R_0 \approx 4.8$.  We present results
for two ``cubed sphere'' grids, one with
$N_R \times N^2$ =$50\times 29^2$ cells in
each of the six sectors
(a total of $\approx 2.5\times10^5$ cells), and
the other with
$26 \times 15^2$ cells per sector (a total
of $\approx 3.5\times  10^4$ cells).
    For the first case
$q=1.055$ while for the lower resolution case
$q=1.11$.

\subsection{Boundary Conditions}

       At the inner boundary
$R=R_{\rm min}$ - the ``numerical star,'' -
we assume  ``free" boundary
conditions
${\partial/\partial R}=0$
for all variables.  
   This boundary condition corresponds
to absorption of incoming matter so
that there is no standoff shock.
      This boundary is treated as a
rotating perfect conductor
${\bf \Omega}=\Omega \hat{\bf z}$.
The flow velocity  ${\bf v}$
was corrected such way that
   in the reference frame rotating with
the star
it is parallel to
${\bf B}$ at $R=R_{\rm min}$,
that is,
${\bf B \times}{\bf v} = {\bf0}$ at $R_{\rm min}$.
       The boundary
condition at $R_{\rm min}$ on the magnetic field
has ${\partial(R B_\phi)}/{\partial R}=0$.
At the outer boundary $R=R_{\rm max}$
   free boundary conditions are taken for all variables.

\subsection{Initial Conditions}

        A star of mass $M$ is located
at the origin of the coordinate system.
      The initial magnetic field is the
tilted dipole field of the star,
${\bf B}_0={3(\rvecmu \cdot {\bf R}){\bf R}}/{R^5}
- {\rvecmu}/{R^3},
$ where $\rvecmu$ is tilted by an angle $\Theta$
from the $z-$axis (which is $\parallel {\bf \Omega}$).
       As initial conditions we set up a
low-temperature  $T_d$,
high-density $\rho_d$ disk
with a high-temperature $T_c\gg T_d$,
low-density $\rho_c \ll \rho_d$
corona filling the remainder of the simulation region.
        The disk rotates with the same axis
as the rotation of the star with
angular velocity close to the
Keplerian value $\omega \approx \Omega_K$.
        The disk extends inward to a radius $R_0$
as discussed by Romanova {\it et al.} (2002)
for the case of an aligned rotator.
   At this distance
the ram pressure of the disk matter is of the order
of the magnetic pressure
of the dipole, $p+\rho {\bf v}^2  = B^2/{8\pi}$.
        Initially, at
any cylindrical radius $r$ from
the rotation axis, we rotate
the corona and the disk
at the same angular rate.
    This avoids a jump
discontinuity of the angular velocity
of the plasma at the boundary
between the disk and the
corona.
       Of course this distribution of $\omega$ in the corona
leads to twisting of the dipole magnetic field
lines.
        However, the twisting
is smoothly distributed along the
     magnetic field lines, and it does not
lead to fast evolution of the disk-corona system.

\subsection{Godunov-Type Finite-Difference Scheme}

    All variables are evaluated at the
centers of the cells.
    All vector variables are expressed
in terms of their Cartesian
components.
     Finite difference
equations are written for Cartesian
components of vector variables.
     The
finite difference scheme of Godunov's
type has a form:
\begin{equation}
{\frac{ {\cal U}^{p+1}-{\cal U}^p}{\Delta t}} V +
\sum_{m=1..6} s_m {\cal F}_m = {\cal Q}~.
\end{equation}
  Here,
${\cal U}=\left \{\rho,~ \rho {\bf v},
 {\bf B}
~\rho S\right \} $  is the ``vector"
of densities of conserved variables;
${\cal F}_m$ is the ``vector" of flux
densities normal to the face ``$m$" of
the cube,
$s_m$ is the area of the face ``$m$",
$V$ is the volume of the cell,  $Q$
is the intensity of sources in the
cell, and $\Delta t$ is the time step.

To calculate the
flux densities ${\cal F}_m$, an approximate
Riemann solver is used, namely, an
eight-wave Roe-type approximate
Riemann solver analogous to one described by
Powell
{\it et al.} (1999).
The  ${\cal F}$'s are represented as
\begin{equation}
{\cal F}_m={1\over2}({\cal F}' + {\cal
F}'') -
\frac{1}{2}\sum_a \nu_a |\lambda_a| A_a {\cal
R}_a~.
\end{equation}
Here, ${\cal F}'$ and ${\cal F}''$
are flux densities normal to the face ``$m$''
of the cell coming from neighboring
cells, and $\nu_a$ are numerical coefficients
which regulate the accuracy of the method. 
   These  are calculated using
values of variables in the cells
which are separated by this face.
    The $\lambda_a$'s are velocities of the
waves in the direction of the normal to the face;
the $A_a$'s are the wave amplitudes; and the
${\cal R}_a$'s are the right-hand
eigenvectors for the eight types of waves,
$a=(E,B,\pm A, \pm F, \pm S)$  (see, e.g., Brio \& Wu 1984;
Powell et al. 1999).

  As a test of our 3D code, spherical 
Bondi accretion (Bondi 1952) was
simulated for the case of a magnetic monopole field, 
${\bf B} \propto 1/R^2$ with a grid $50 \times 11^2$ in
each sector.
  The simulations  shown good agreement with
the radial dependence of the Bondi solution
and with the predicted accretion rate.

\section{Simulations of Funnel Flows}

   Here, we discuss  simulation
results for the case where the star's magnetic
moment is inclined at an angle $\Theta=30^\circ$
to the rotation axis ${\bf \Omega}$ which is parallel
to the normal of the disk.

       The magnetic
moment of the star and the density of
the disk are such that the
ram pressure of the disk is
approximately equal to the
   magnetic pressure of the dipole at
the inner radius of the disk $R_0$ at
$t=0$.
    Thus, the funnel flow (FF) is
expected to start from about
this radius, as  observed in
our axisymmetric simulations
(Romanova {\it et al.} 2002).
   We rotate the star with angular
velocity $\Omega_* = g (GM/R_0^3)^{1/2}$, with $g=0.19$.
      The corotation radius  of the star $r_{cor}$
is distance where the centrifugal force
$\Omega_*^2 r$ equals the gravitational
force $GM/r^2$;  that is,
$r_{cor}=(GM/\Omega_*^2)^{1/3}$.
    It follows that $r_{cor}=R_0/g^{2/3} \approx 3 R_0$.

     Some information about the
$3D$ MHD flow can be obtained from the
three orthogonal slices, an $(x,z)$ slice at $y=0$,
  an $(y,z)$ slice at $x=0$, and an $(x,y)$ slice
at $z=0$.
 Figure 2 shows these
cross-section for different grids.
    The most refined grid used in the present
simulations has
$N_R\times N^2 $ = $50\times
29^2$ cells in each of the
six sectors of the cubed sphere.
  The coarsest grid has
$26\times 15^2$ cells in each sector.
     Linking with a spherical
coordinate system $(R,\theta,\phi)$,
we have the correspondence for the
number of cells,
  $N_R\times N_\theta\times N_\phi=
6 N_R\times  N^2$.
      We observed that the slices
of the accretion flows are closely
similar for the different grids.
    Namely, with decreasing distance,
the disk ends and
matter starts to flow out of the
plane of the disk into the funnel flow
at radii $R\approx 2R_0$ in $x-$direction
and at $R\approx 1.7R_0$ in $y-$direction.
   In the $(x,z)$ slice (which is the plane containing
$\rvecmu$ and ${\bf \Omega}$) the funnel flow
takes the longer of the two paths along
the magnetic field.
    This we find to be a distinctive effect
of the three-dimensional accretion flow
to a mis-aligned dipole.
    Figure 3 shows a three dimensional view of
the funnel flow.   Close to the star the flow
is in two streams which approach the star
from opposite sides.

    As a test of our code,
we did $3D$ simulations of accretion
to an aligned rotator ($\rvecmu \parallel {\bf\Omega}$) and
  analogous $2D$ axisymmetric simulations in
spherical coordinates with the number of cells
in the meridional plane
$N_R\times N_\theta = 50\times 29$.
    Results of the $2D$ and $3D$ simulations
show qualitatively similar funnel flows
at the same elapsed times.

\section{Conclusions}

   We developed a new three
dimensional MHD simulation code for studying
the accretion flows to rotating stars with
mis-aligned dipole magnetic fields.
   New three-dimensional features are found
in the  simulations of disk accretion
to a rotating star with dipole moment $\rvecmu$
inclined by an angle $\Theta$ to
the star's rotation axis.
    Specifically, in
the $x-z$ cross section of the flow containing
$\rvecmu$ and ${\bf \Omega}$, the funnel flow takes
the longer of the two possible paths along magnetic
field lines to the surface of the star.
     Furthermore, the funnel
flow to the stellar surface  is mainly in  two streams
which approach the star from opposite directions.
A subsequent paper will give detailed analysis of
the funnel flows at different inclination angles $\Theta$.

\acknowledgments
This work was supported in part by NASA grants
NAG5-9047 and NAG5-9735 and by NSF grant AST-9986936.
 RMM grateful to NSF POWRE grant for partial support.
AVK and GVU were partially supported by INTAS
grant 00-01-00392,
and RFBR grant 00-02-17253.

\end{document}